\documentclass{aa}  

\usepackage{graphicx}
\usepackage{soul}
\usepackage[normalem]{ulem}
\usepackage{txfonts}
\begin{document}

   \title{
   Unveiling the magnetic nature of chromospheric vortices}


   \author{Mariarita Murabito,
          \inst{1}
          Juie Shetye,
          \inst{2,3}
          Marco Stangalini,
          \inst{4}
          Erwin Verwichte,
          \inst{2}
          Tony Arber,
          \inst{2}
          Ilaria Ermolli,
          \inst{1}
          Fabrizio Giorgi,
          \inst{1}
          Tom Goffrey
          \inst{2}
          }

   \institute{INAF Istituto Nazionale di Astrofisica, Osservatorio Astronomico di Roma, 00078 Monte Porzio Catone, RM, Italy \\
              \email{mariarita.murabito@inaf.it}
         \and
                The Centre for Fusion, Space and Astrophysics, Department of Physics, University of Warwick, Coventry CV4 7AL, UK 
        \and
                Armagh Observatory and Planetarium, College Hill, Armagh BT61 9DB. 
        \and
                ASI - Agenzia Spaziale Italiana, Via del Politecnico snc, Rome, Italy}
            

   \date{Received XXXX; accepted XXXX}

 
  \abstract
   {Vortex structures in the Sun's chromosphere are believed to channel energy between different layers of the solar atmosphere. 
   }
   {We investigate the nature and dynamics of two small-scale quiet-Sun rotating structures in the chromosphere.}
   {We analyse two chromospheric structures that show clear rotational patterns in spectropolarimetric observations taken with the Interferometric Bidimensional Spectrometer (IBIS) at the Ca II 8542 \AA~ line.
   }
   {We present the detection of spectropolarimetric signals that manifest the magnetic nature of rotating structures in the chromosphere.
  Our observations show two long-lived structures of plasma that each rotate clockwise inside a 10 arcsec$^{2}$~ quiet-Sun region. Their circular polarization signals are 5-10 times above the noise level. Line-of-sight Doppler velocity and horizontal velocity maps from the observations reveal clear plasma flows at and around the two structures. An MHD simulation shows these two structures are plausibly magnetically connected. Wave analysis suggests that the observed rotational vortex pattern could be due to a combination of slow actual rotation and a faster azimuthal phase speed pattern of a magneto-acoustic mode.}
   {Our results imply that the vortex structures observed in the Sun's chromosphere are magnetic in nature and that they can be connected locally through the chromosphere.}

   \titlerunning{Unveiling the magnetic nature of chromospheric vortices}
   \authorrunning{Murabito M. et al.}
   \maketitle

%

\section{Introduction}

Rotating structures have been observed in both active- and quiet-Sun regions of the solar atmosphere, and on a variety of spatial and temporal scales, since the beginning of the 20th century  \citep[see for instance][]{Haleb}. 
Terminologies such as vortices, tornados or swirls have been employed to identify such rotating structures. For the sake of consistency, we shall call these structures vortices.
Using modern telescopes supported with adaptive optics systems, vortices of few hundred kilometers in size have been identified across the solar atmosphere \citep[e.g.,][]{Bonet2010,Attie2009,Vargas2011,Wedemeyer2012,Chroswirls,Tziotziou2018}. Quiet-Sun photospheric vortices have a typical size of 0.5-2 Mm and a mean lifetime up to 15 minutes \citep{Bonet2010,Wedemeyer2013,Liu2019b}.
 There are estimated to be 3.1$\times$10$^{-3} $ photospheric vortices Mm$^{-2}$ minute$^{-1}$ \citep{Bonet2010,Vargas2011}. Photospheric vortex flows are usually correlated with network magnetic elements at supergranular vertices \citep{Requerey2018}. In the quiet-Sun chromosphere, vortices have typical sizes of 1.5-4.4 Mm and a lifetime of about $10-15$  minutes up to 1.7 hr \citep{Chroswirls,Tziotziou2018}. They are seen to rotate as quasi-rigid bodies \citep{Wedemeyer2012,Tziotziou2018}.
 
Numerical simulations have revealed that photospheric rotation forces the overlying plasma to rotate and lift up, and the magnetic field to twist \citep{Wedemeyer2012,Wedemeyer2014,Yadav2020}. 
Also, simulations show that rapidly changing photospheric perturbations propagate vertically along the magnetic field at the local Alfv\'en speed \citep{Wedemeyer2012,Shelyag2013,Liu2019c}. It has been proposed that vortices are Alfv\'{e}n waves that carry an average of 100-300 W m$^{-2}$ Poynting flux into the upper atmosphere. These results support the findings on the quasi-periodic oscillations and short lifetimes of photospheric magnetic vortices reported from the simulations performed by \citet{Moll2011} and \citet{Kitiashvili2013}. 
Vortices could also generate shocks and support magnetoacoustic waves \citep[e.g.][]{Fedun2011,Shelyag2013,Tziotziou2019}. \cite{Shetye2019} have shown that chromospheric acoustic waves interact with vortices, leading to modified wave characteristics that may be used to probe vortex structure. 

These studies are indicative of a magnetic nature of chromospheric vortices. 
However, direct observational evidence of the magnetic field of vortices in the upper layers of the solar atmosphere are still lacking. Extracting magnetic properties in the quiet-Sun chromosphere remains a challenge.  
 Nevertheless, from the available chromospheric lines, the Ca II infrared triplet 
 constitutes one of the best candidates for the analysis of chromospheric magnetic field \citep{QuinteroNoda16}. In this paper, we present the first spectropolarimetric measurements of two chromospheric rotating structures.

\section{Observational signatures}

We analyse observations acquired with the Interferometric Bidimensional Spectrometer \citep[IBIS,][]{Cav06} of a quiet-Sun region on May 1, 2015, from 14:18 UT to 15:03 UT, near disk center. They consist of spectropolarimetric observations at a 46s time cadence using photospheric Fe I 6173 \AA~and chromospheric Ca II 8542 \AA~lines sampled in twenty-one spectral points 
with transmission of 20 m\AA~and 60 m\AA, respectively. 

The field of view (FOV) is 40\arcsec$\times$80\arcsec~with a pixel scale of 0.08\arcsec.
Broadband white-light images with the same time cadence and FOV are taken at 6333.20 $\pm$ 0.05 \AA.  
In addition to the standard calibration procedures,

the data were restored with the Multi-frame Blind Deconvolution \citep[MOMFBD,][]{Lof02}. Although the polarization signals detected in the chromosphere are low, the observations were taken under exceptional seeing conditions for the entire duration of the data sequence. This is also confirmed by the fact that the AO system was able to lock on the quiet-Sun, a challenging task for solar wavefront sensing. The exceptional conditions during observations render this dataset unique for data exploitation.   
We focus on a 11.3\arcsec$\times$12\arcsec\ region of interest (ROI) that includes two rotating structures.


\begin{figure}
  \centering
    \includegraphics[scale=0.4, clip, trim=185 50 250 50]{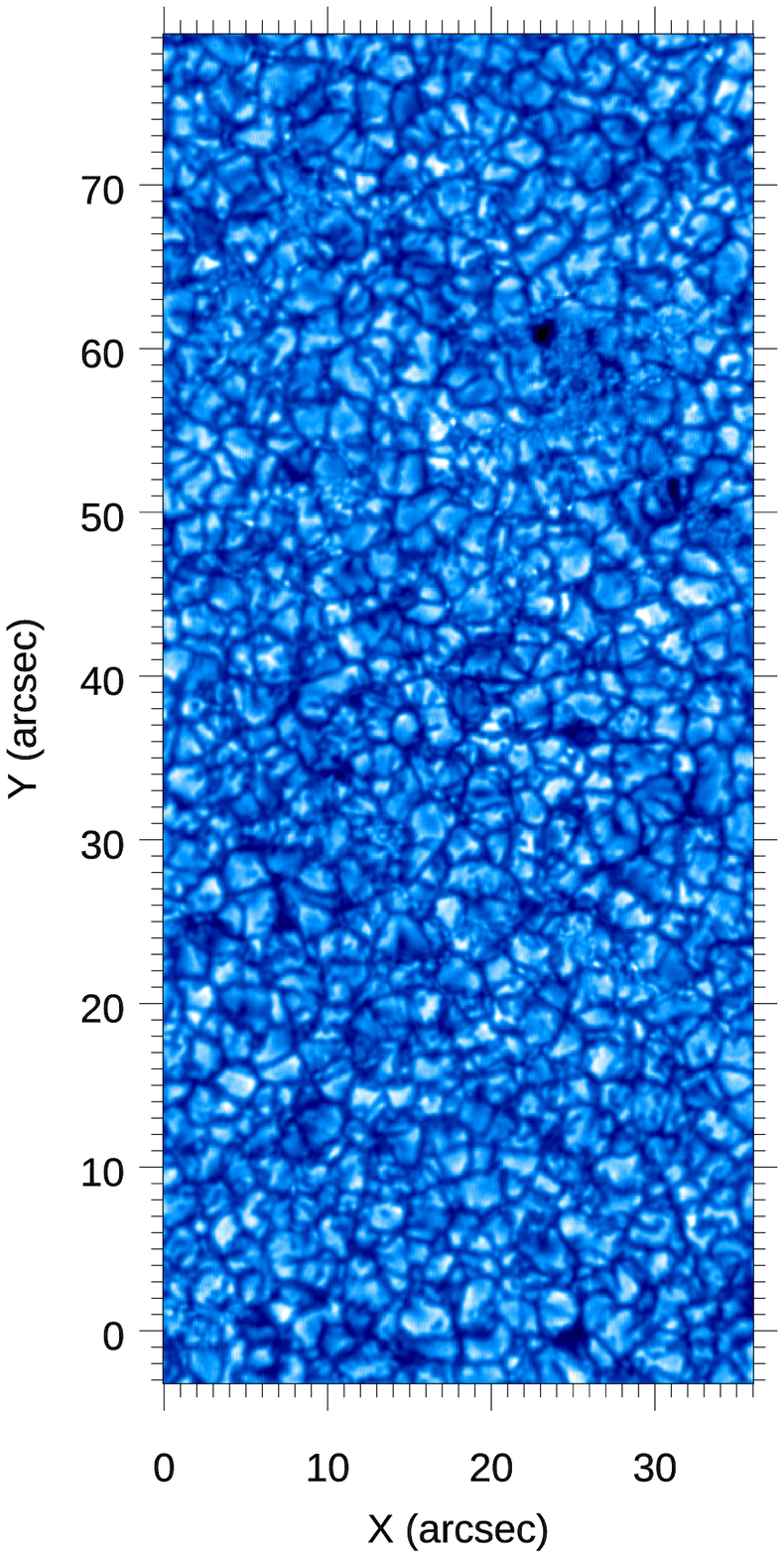}
    \includegraphics[scale=0.4, clip, trim=180 50 240 50]{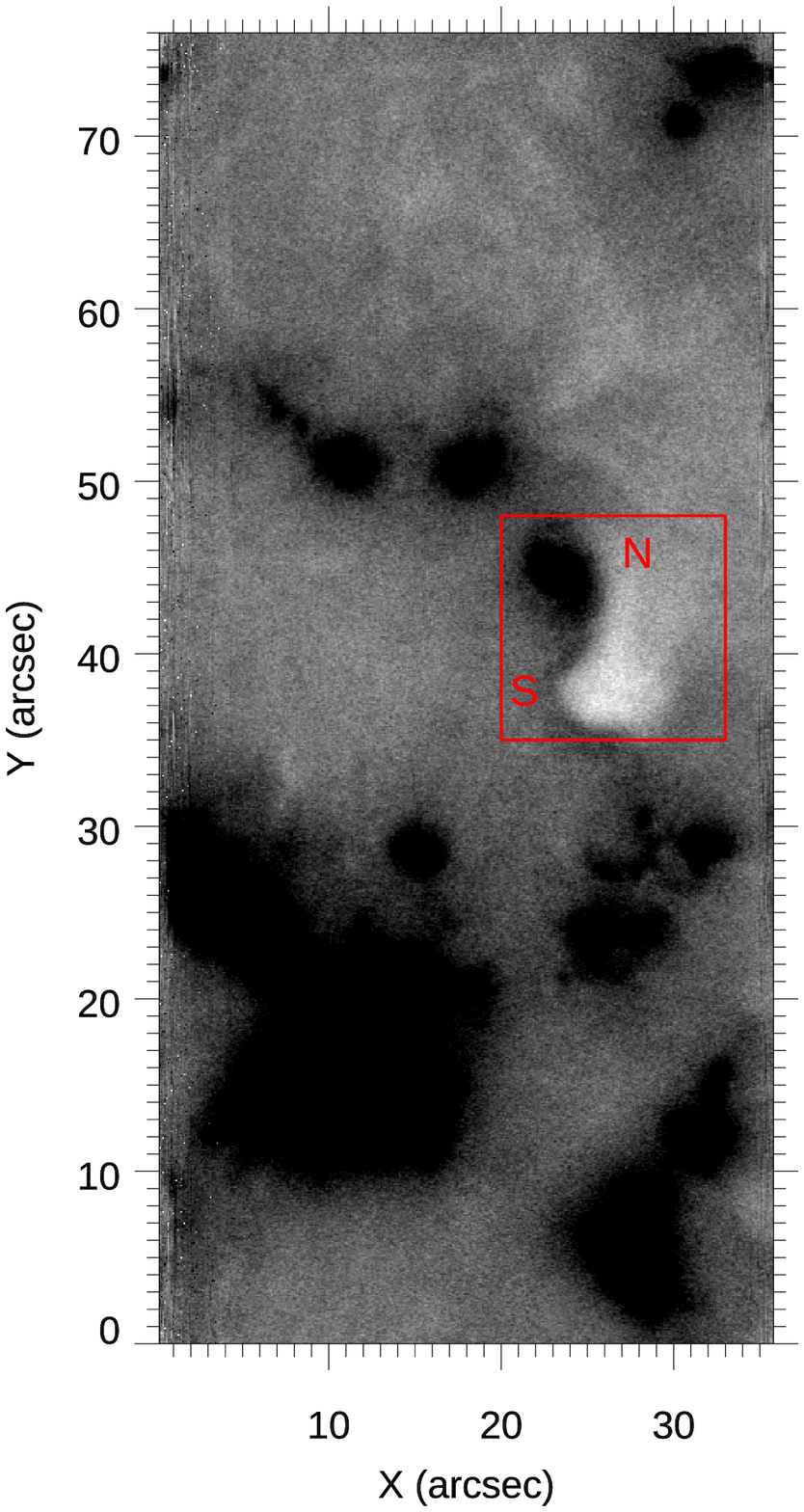}
    \includegraphics[scale=0.98, clip, trim=0 0 10 0]{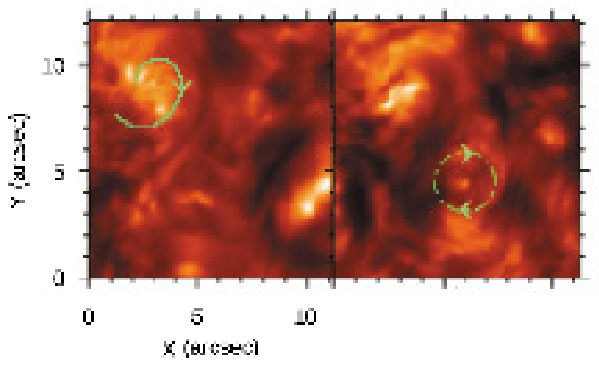}
    \caption{
    Top left: White-light image at 6333 \AA~of the IBIS FOV at 14:20 UT.
    Top right: Mean CP map from IBIS Ca II 8542 \AA~data. The signal in the CP map has been saturated to enhance contrast. Bottom: Ca II core snapshots with curves to emphasise the rotation pattern in the two structures, following the classification by \citet{Chroswirls}.
 \label{fig:fig1}}
\end{figure}

Figure \ref{fig:fig1} shows the white-light image and the map of mean chromospheric circular polarization (CP) of the IBIS full FOV. When saturated to enhance relevant features, a clear pairing of opposite CP signals is revealed in the ROI.
CP is a measure of the line-of-sight (LOS) magnetic field, while the linear polarization (LP) of the inclined magnetic fields. We computed the CP and LP signals pixel-wise following the method set out by \cite{Pillet2011} and adapted to our specific data set as

\begin{equation}
\left[\!\!
\begin{array}{c}
\mathrm{LP} \\ \mathrm{CP}
\end{array}
\!\!\right]
\,=\, 
\frac{1}{10 \left \langle I_{c} \right \rangle} \sum_{i=5}^{15}
\,\left[\!\!\!\!
\begin{array}{c}
\sqrt{Q^{2}_{i}+U^{2}_{i}} \\ \epsilon_ {i}  |{V_{i}}|
\end{array}
\!\!\right]
\,\,\,,\,\,
\epsilon_i \,=\, \left\{
\begin{array}{lr}
1 & i < 10 \\
0 & i = 10 \\
-1 & i > 10
\end{array}
\right.
,
\end{equation}

where $I_i$, $Q_i$, $U_i$ and $V_i$ are the Stokes parameters at line position $i$ and $\left \langle I_{c} \right \rangle$ is the average Stokes $I$ in the line-continuum in a quiet region within the FOV.

\begin{figure}
   \centering
   \includegraphics[scale=0.4]{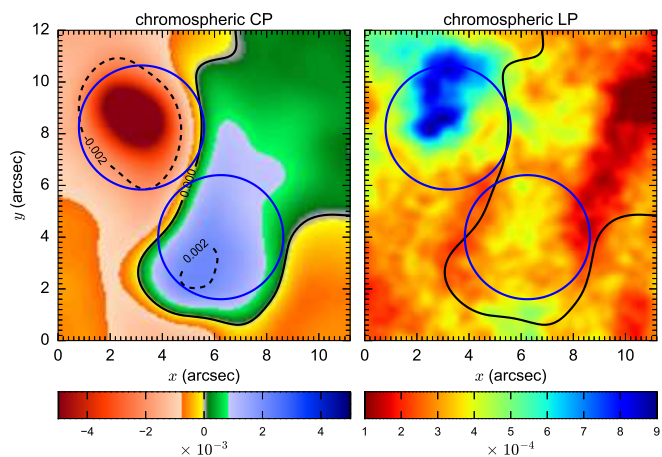}
    \caption{Time-average CP (left) and LP maps (right). Contours indicate the polarity inversion line (solid line) and values of $\pm$2$\times$10$^{-3}$ (dashed lines). Blue circles are centred on the N and S structures.
 \label{fig:cp_lp}}
\end{figure}

 Figure \ref{fig:cp_lp}, left panel, displays the time-average CP map which indicates the presence of a magnetic dipole in the chromosphere separated by a curved polarity inversion line. 
Unfortunately, the $U$ and $Q$ Stokes signals that enter the LP calculation are too weak to extract meaningful information at each time step. Instead, we integrate the signal in time and produce an average LP map (see right panel of Figure \ref{fig:cp_lp}). 
It is likely that there are a horizontal components of the magnetic field in and around the studied structures.

The movie available in the online material shows two structures at the location of this dipole. They are present during the entire observation period. For simplicity, we name the two structures North (N) and South (S). They consist of small-scale bright and dark patches forming a pattern of spiral arms and concentric rings, showing clockwise rotation. In particular, following the schematic classification 
introduced by \citet{Chroswirls}, we identify two different patterns in the Ca II line core maps (see bottom panel of Figure \ref{fig:fig1}). The S structure exhibits a roughly concentric ring pattern with a radius of about 1.5\arcsec~during most of the observational time, while the N structure shows spiral arms pattern.

The patches in S are clearest and rotate faster than in N. For this reason, we identify S as a vortex, but refer to N as a rotating 'structure'. The CP and LP signal (shown in Figure \ref{fig:cp_lp}) represent the first chromospheric spectropolarimetric measurement of a vortex. There is also a dark feature moving from the south side of N to the east side of S, along a curved trajectory. It wraps around S in the opposite sense than S rotation. There is a degree of uncertainty in flow tracking as seeing conditions, though excellent, may have affected contrast in regions of darkest intensity.

\begin{figure*}
    \centering
    \includegraphics[scale=0.65,clip, trim=0 80 0 370]{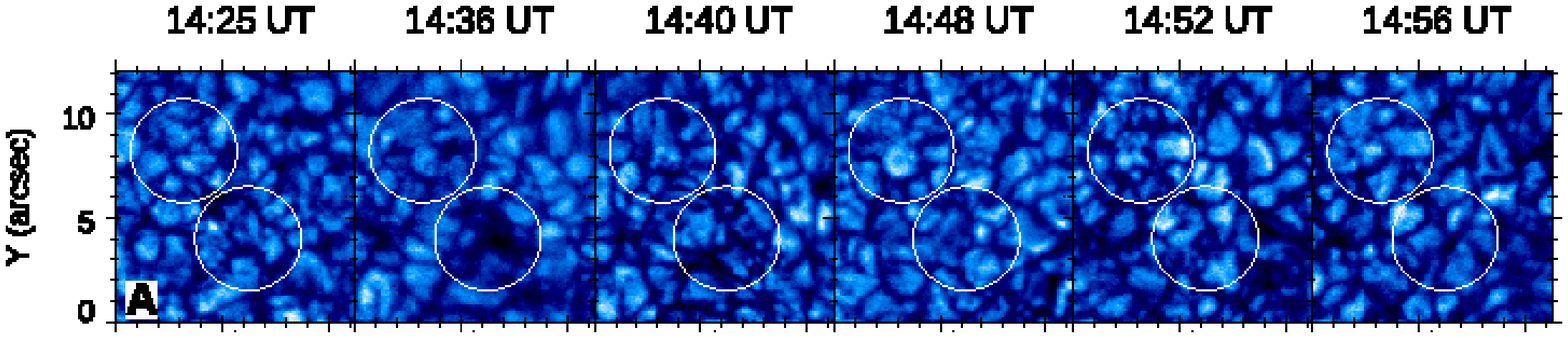}
 \includegraphics[scale=0.65,clip, trim=0 80 0 380]{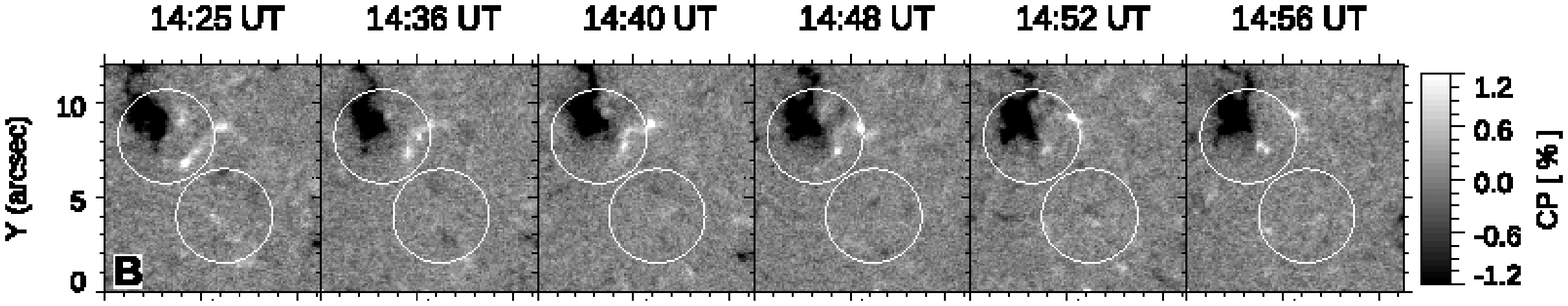}
 \includegraphics[scale=0.65,clip, trim=0 280 0 180]{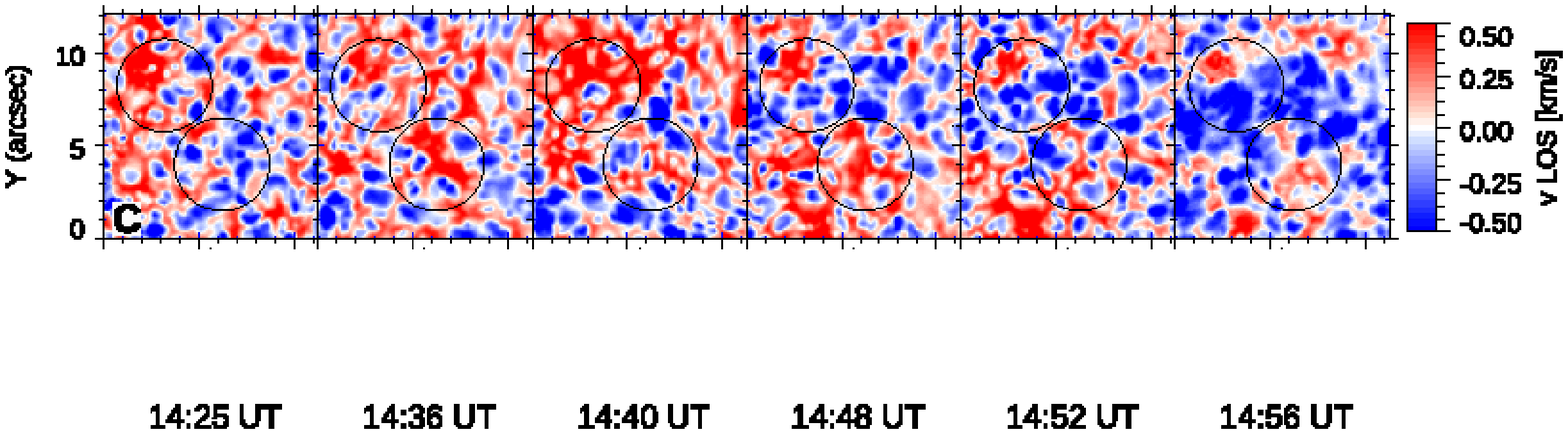}
 \includegraphics[scale=0.65,clip, trim=0 80 0 380]{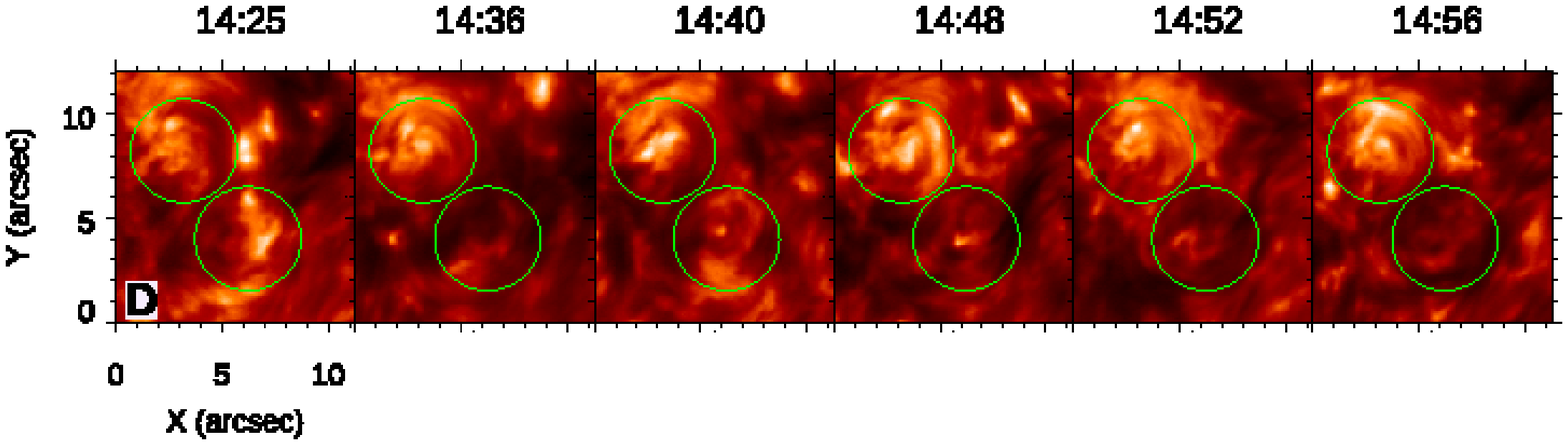}
  \includegraphics[scale=0.65,clip, trim=0 280 0 180]{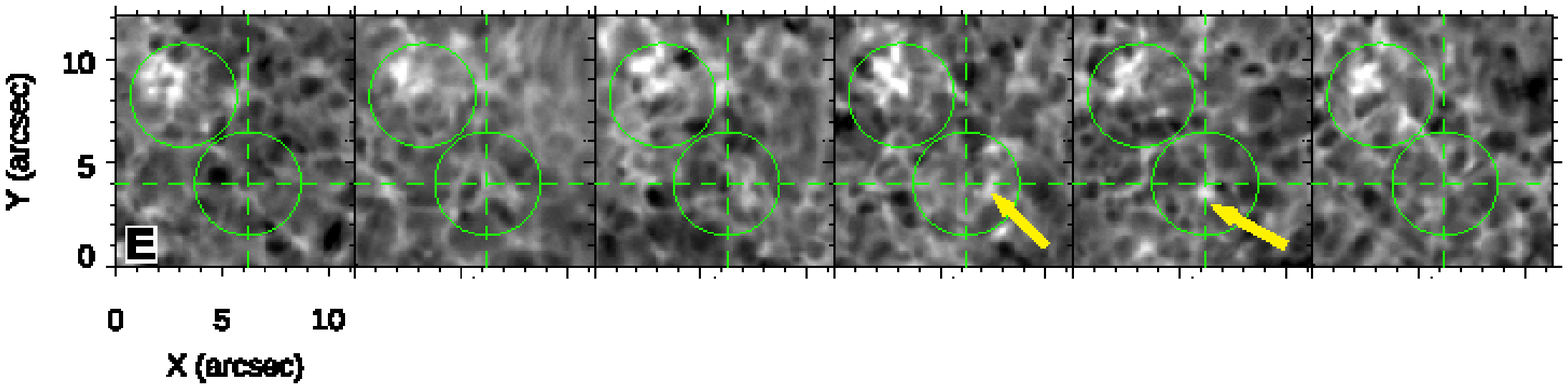}
\includegraphics[scale=0.65,clip, trim=0 80 0 380]{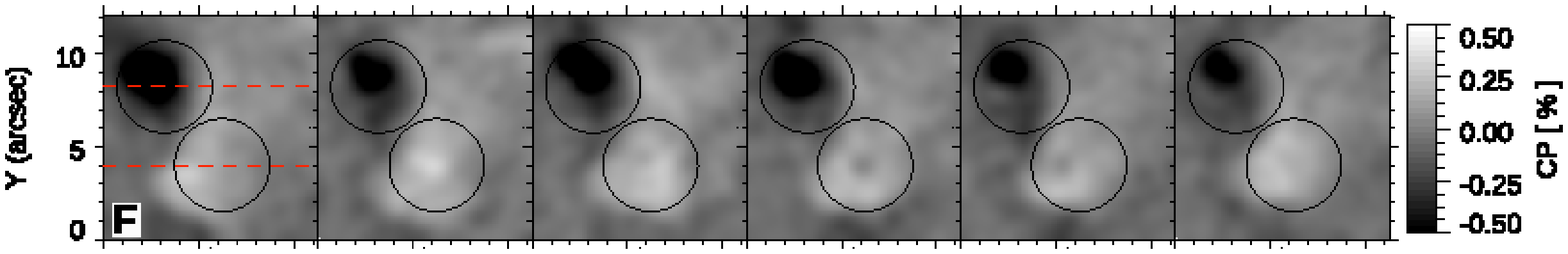}
\includegraphics[scale=0.65,clip, trim=0 230 0 180]{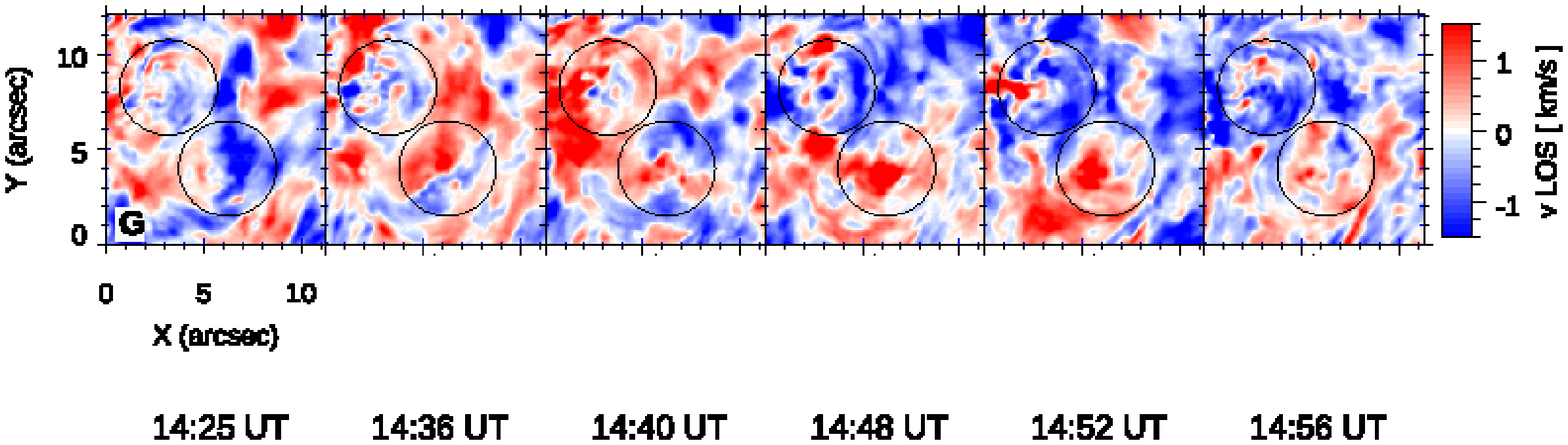}
    \caption{Panels A,B and C: continuum intensity, CP maps and LOS filtered velocity map from Fe I line measurements. Panels D,E, F and G: line core intensity, line blue wing, CP maps and LOS velocity maps from Ca II line measurements. 
    Down--flow and up--flow correspond to positive and negative velocities, respectively. A Gaussian filter is applied to the chromospheric CP maps (panel F). The dashed horizontal lines indicate the slices where the time slice is computed (see Section 2 for more details). (A movie of the data presented in this figure is available in the online material.)   \label{fig:3bis} }
\end{figure*}

Figure~\ref{fig:3bis} shows in more detail the evolution of the ROI in the photosphere (panels A, B and C) and in the chromosphere (panels D, E and F), respectively.
At the photosphere, intensity images and CP maps show that the N structure is anchored in an extended concentration of strong negative polarity magnetic flux. The CP signal in the N structure is larger than 1$\%$ (the noise level of the measurements is around 0.01$\%$). The S vortex lies above a patch of positive, spatially diffused, magnetic flux. Unusually for a solar vortex, the S structure cannot be associated with a distinct photospheric flux concentration. Instead, it corresponds to a darker extended intergranular region.


The LOS velocity maps shown in Figure~\ref{fig:3bis} (panels C and E), have been constructed pixel-wise using a fit of Stokes I with a linear background and a Gaussian line profile. The obtained velocities are band pass filtered to remove p-modes. 
The photospheric LOS velocity map (see panels C) looks similar to the one expected from a pattern of
granules, even if at 14:25 UT a strong red--shifted patch appears at the position of N, showing a down-flow at the magnetic structure. 

The Ca II line core images (see panels D in Figure~\ref{fig:3bis}) show 
two structures at the location of each of the two opposite polarity flux concentrations consisting of bright and dark regions. 
Observations of the middle-upper photosphere sampled by the Ca line wing (see panels E) display small brightening events near the centre of the S structure (pointed by the yellow arrow). There are no such small events at the N structure. Instead, we detect a larger brightening region associated with the negative photospheric magnetic concentration. 

Chromospheric CP maps (see panels F in Figure~\ref{fig:3bis}) display a compact negative patch (with values larger than 0.5\% and up to 1.5\%) co-spatial with N and a more diffused opposite polarity patch (with values up to 0.9$\%$) at S. The N structure displays a mixture of up- and down-flow regions of the order of 0.5-0.7 km s$^{-1}$ in chromosphere (see panels G in Figure~\ref{fig:3bis}). The Eastern side shows up-flows of about 0.5 km/s. At the centre of the S vortex is a persistent down-flow (panels from 14:36 UT to 14:52 UT) of 0.5$\pm$0.3 km s$^{-1}$.


\begin{figure}
    \includegraphics[scale=0.52,clip, trim=20 370 60 200]{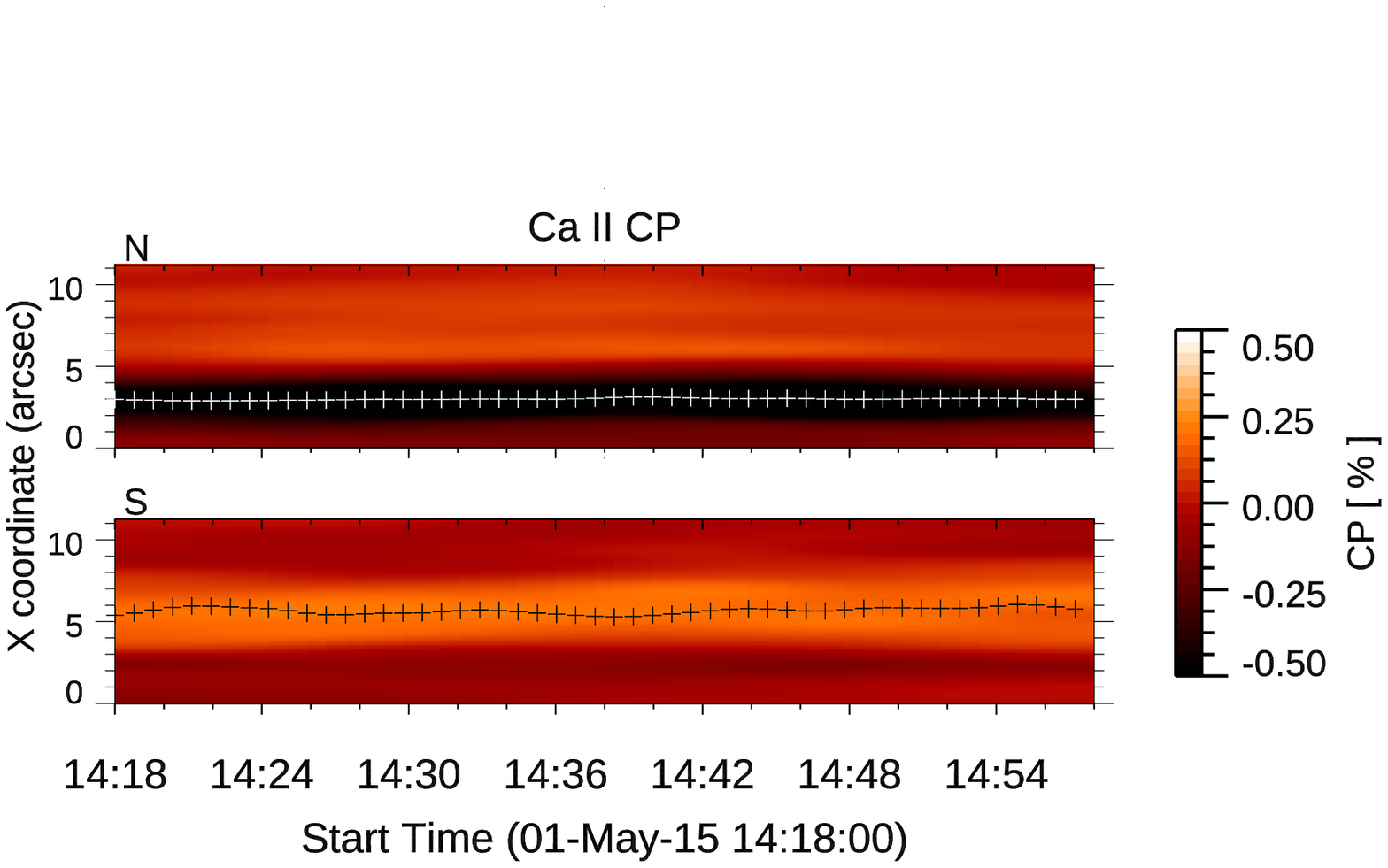}
    \caption{Time-distance plot of the chromospheric circular polarization along a cut at the centre of the two studied structures. The $x$-position of the baricenter of each structure is overplotted with a plus symbol. A Gaussian filter is applied to these plots. \label{fig:cp} }
\end{figure}

Figure \ref{fig:cp} shows time-distance plots of the CP signal along the two profiles indicated in the first panel of CP in Figure \ref{fig:3bis}. To better highlight the two structures, the $x$-position of their barycentre is annotated. These values were computed identifying each structure with a threshold in the CP. Both the time-slice plots show that these structures are magnetically stable in time with a value of CP more than 0.5$\%$ and 0.2$\%$, for the N and S structures, respectively. It is noteworthy that the time-slice plot centered on S exhibits a negative polarity filament at around X=3\arcsec (opposite to the positive polarity at the S vortex). 

\subsection{Rotation} \label{sec:rotation}

\begin{figure}
    \centering
    \includegraphics[scale=0.44,clip, trim=0 0 0 0]{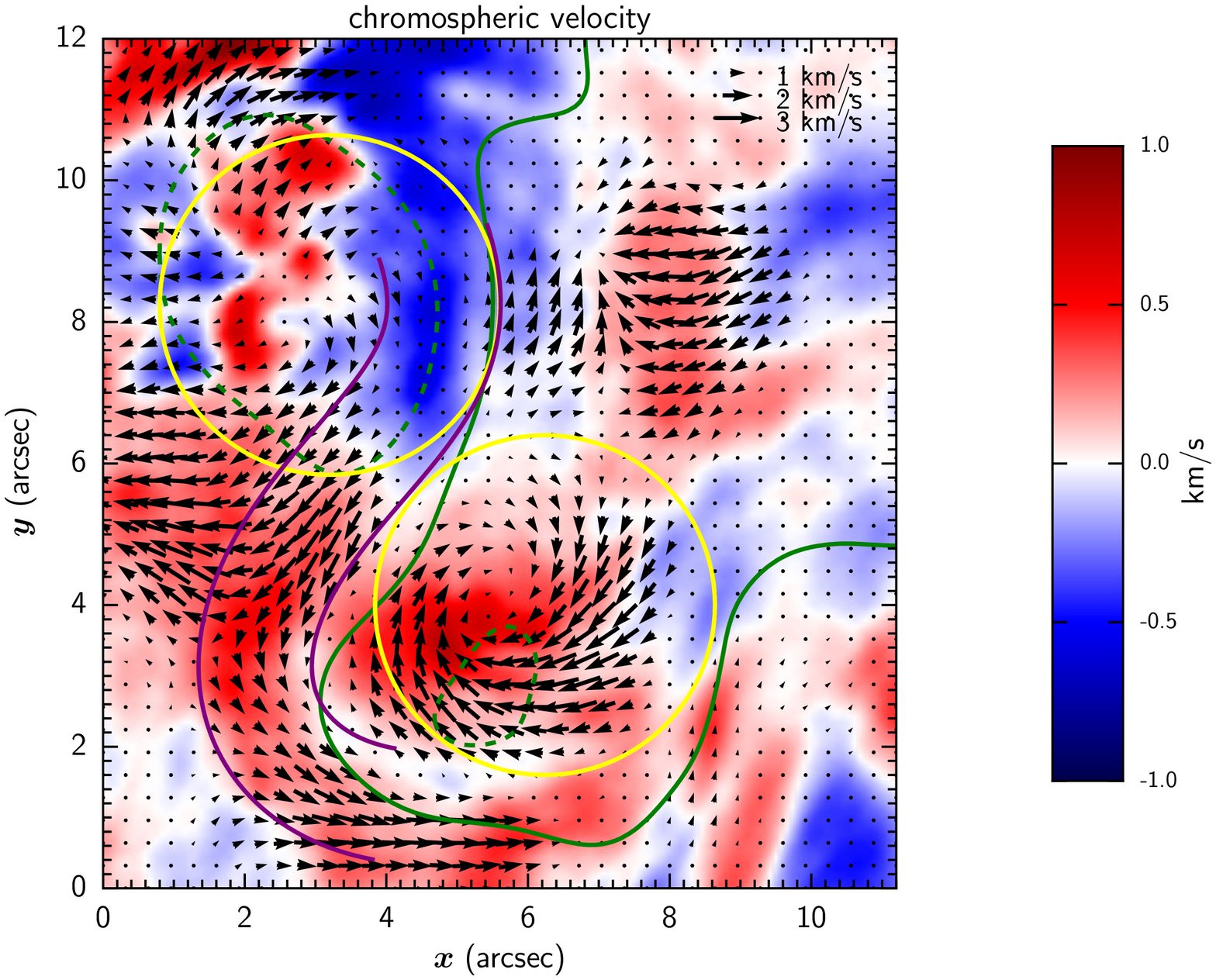}
    \caption{Time-average chromospheric LOS velocity map with superimposed horizontal flow field. Green lines are contours of CP. The two purple lines outline the flow channel from N to S. 
    \label{fig:erwin1} }
\end{figure}

To highlight the rotation of these structures in the chromosphere, we estimate the horizontal velocity field by applying to the Ca II line core data an optical flow method that uses the two-frame motion estimation algorithm by \citet{Farneback} to obtain the displacement field at each time step. Image smoothing is employed to increase the robustness of the displacement field against noise, at the cost of losing some spatial resolution. The average horizontal velocity field is derived from the mean of the displacement field across the whole observation period divided by the sequence time cadence. Figure~\ref{fig:erwin1} displays the mean chromospheric horizontal velocity map together with the LOS velocity. Both structures show a net clockwise horizontal rotation pattern. The S vortex’s rotation pattern is the most obvious and stretches across 2 Mm. 

We map the data onto a polar grid centred on each structure. This enables us to investigate each structure's local properties as a function of radial distance $r$ from its centre and as a function of azimuthal angle $\phi$.

\begin{figure}
   \centering
    \includegraphics[scale=0.45]{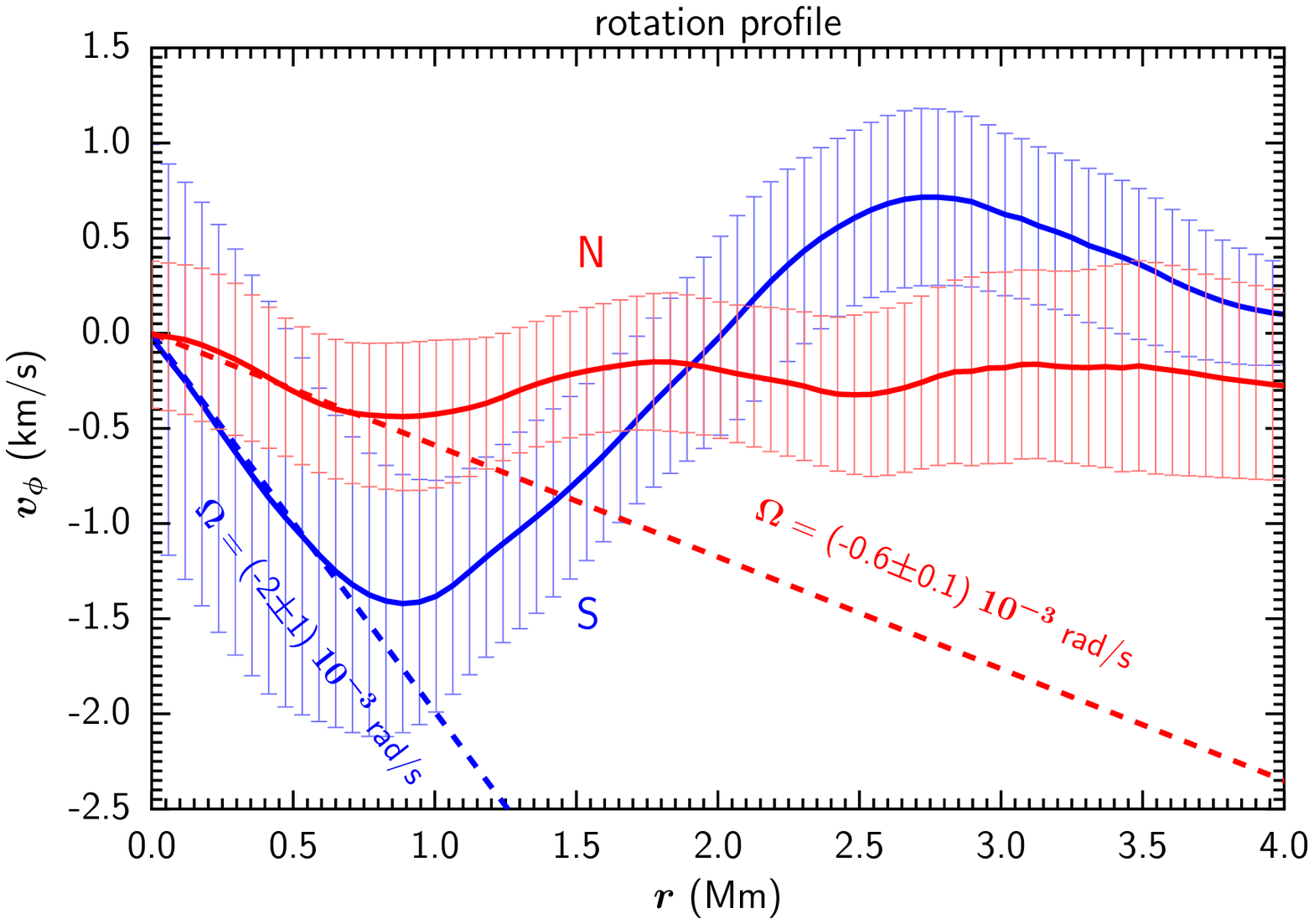}
    \caption{Azimuthal (rotation) speed $v_\phi$ (solid lines) as a function of radius derived from the horizontal velocity in the polar grid centred on each structure averaged in time and over azimuthal angle. The dashed line is a linear fit to the first 0.7 Mm of the profile. \label{fig:rotationprofiles}}
\end{figure}

Figure \ref{fig:rotationprofiles} shows the rotation profiles as a function of radius from the centre of each structure. Near the centre the azimuthal speed $v_\phi$ is approximately negative and linear. This indicates rigid-body clockwise rotation with a constant negative angular velocity. The N structure rotates within a radius of 0.6 Mm at an angular speed of (-0.6$\pm$0.1) $\times$10$^{-3}$ rad/s (period of 3 hours). At larger radii $v_\phi$ remain negative but there is no discernible structure and there may not be a coherent rotation.

The S rotation profile closely matches a rigid body rotation within a radius of about 1 Mm, with an angular speed of (-2$\pm$1) $\times$10$^{-3}$ rad/s (50 minutes). The azimuthal speed reaches a minimum value of -1.5 km/s at $r$=0.9 Mm. 
From $r$=2 Mm the sense of rotation reverses and becomes anti-clockwise. 
This reflects the horizontal flow pattern visible in Figure \ref{fig:erwin1} that surrounds the S vortex to the east and south. There the LOS velocity has a similar magnitude and points downwards.

There is an extended region of negative LOS velocity at the western edge of N (see Figure \ref{fig:erwin1}). This suggests that there is a connecting flow channel between the two structures, i.e. an upflow emanating from near the western edge of N that then proceeds as a horizontal flow towards S. Around the S vortex the flow becomes downwards.

\subsection{Waves} \label{sec:wave}

\begin{figure}
    \centering
     \includegraphics[scale=0.37,clip, trim=0 0 0 0]{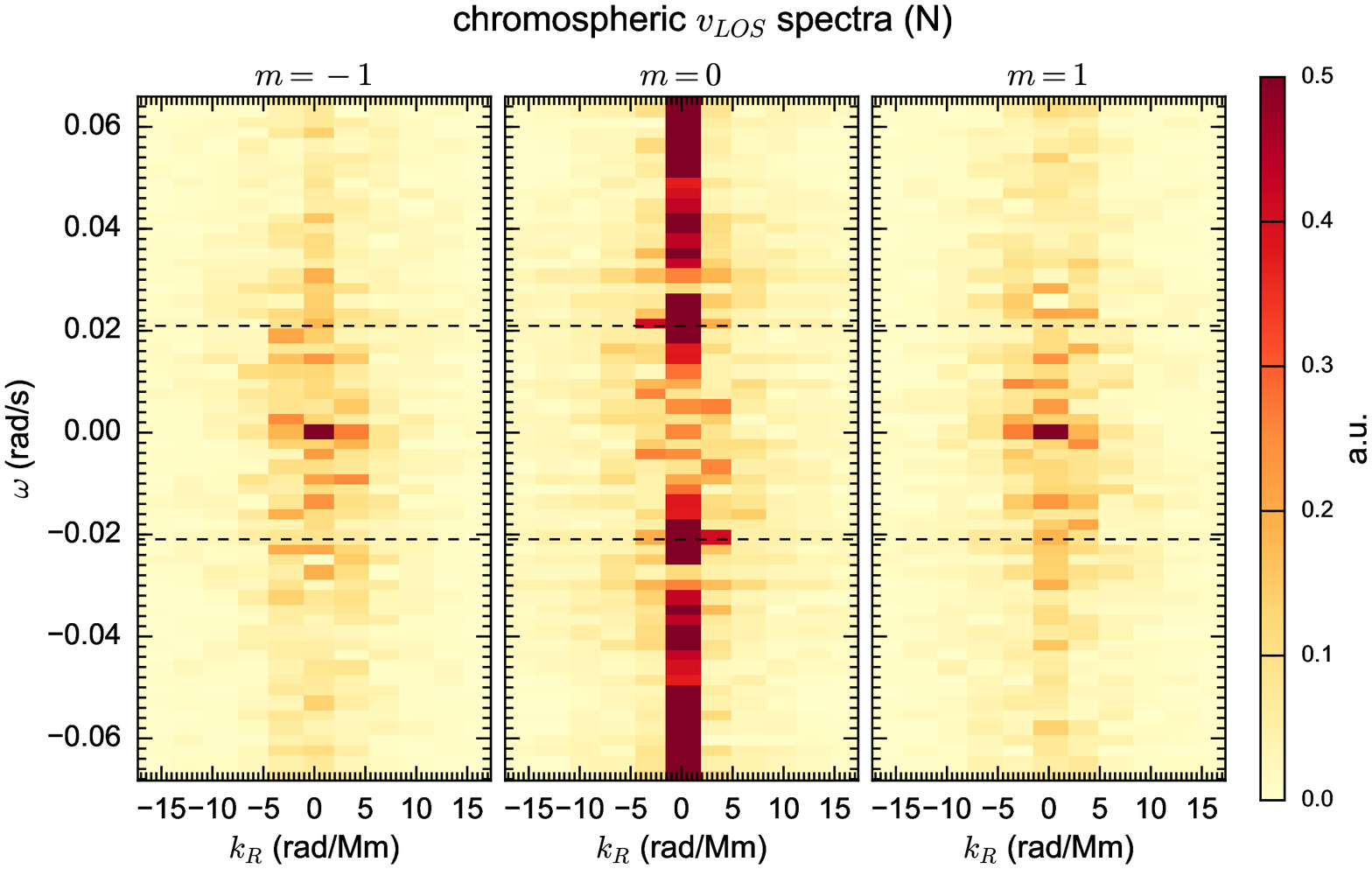}
    \includegraphics[scale=0.37,clip, trim=0 0 0 0]{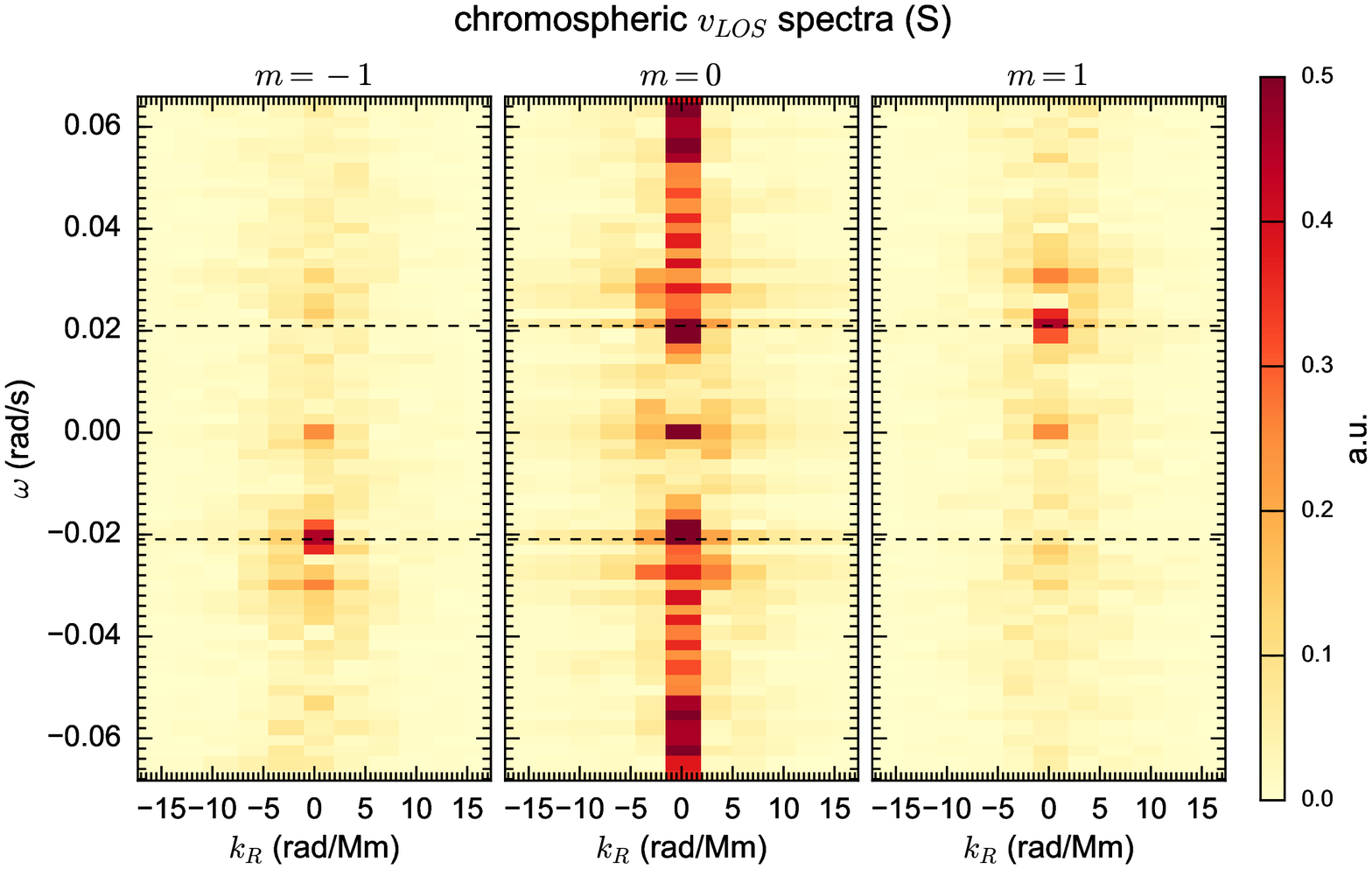}
    \caption{Spectrum of N (top) and S (bottom) polar map in chromospheric LOS velocity as a function of radial wave number $k_R$ and angular frequency $\omega$, for azimuthal wave number $m$=-1 (left),0 (middle),1 (right). The dashed lines indicate a signal with a five minute periodicity. 
    \label{fig:spectraS} }

\end{figure}

The mean velocity field rotation from optical flow is slower than visual estimates of rotation in Ca II intensity.  We examine the dynamics as revealed by the spectrum of the polar maps in terms of radial and azimuthal wave numbers $k_R$ and $m$, and angular frequency $\omega$ for a mode signal of the form $\exp[i(k_R r + m \phi + \omega t)]$. The largest radius of the polar grids used are shown as circles in Figure~\ref{fig:erwin1}.  Figure \ref{fig:spectraS} shows the spectrum for the LOS velocity. The spectra for the N structure show oscillations at angular frequency $\omega_0$=$\pm$0.021 rad/s $m$=0 and $k_R$=3.5 rad/s. This corresponds to an azimuthally symmetric mode of the form $\cos(k_R r - \omega t)$ that propagates radially outwards at a horizontal phase speed of 6.5 km s$^{-1}$ with a periodicity of approximately five minutes. The spectra for the S vortex shows two clear peaks at five minutes with a mode symmetry of $\omega$=-$\omega_0$, $m$=-1 and $\omega$=+$\omega_0$, $m$=+1, which when combined, is a mode of the form $\cos(m\phi + \omega_0 t)$. This represents a five-minute oscillatory mode in the LOS velocity that propagates in the plane of the sky strictly clockwise. It has an $|m|$=1 azimuthal structure, i.e. at any given time each half of the vortex has a signal of opposite sign. A similar five-minute signal is present in intensity and CP.

\begin{figure}
    \centering
    \includegraphics[scale=0.42,clip, trim=0 0 0 0]{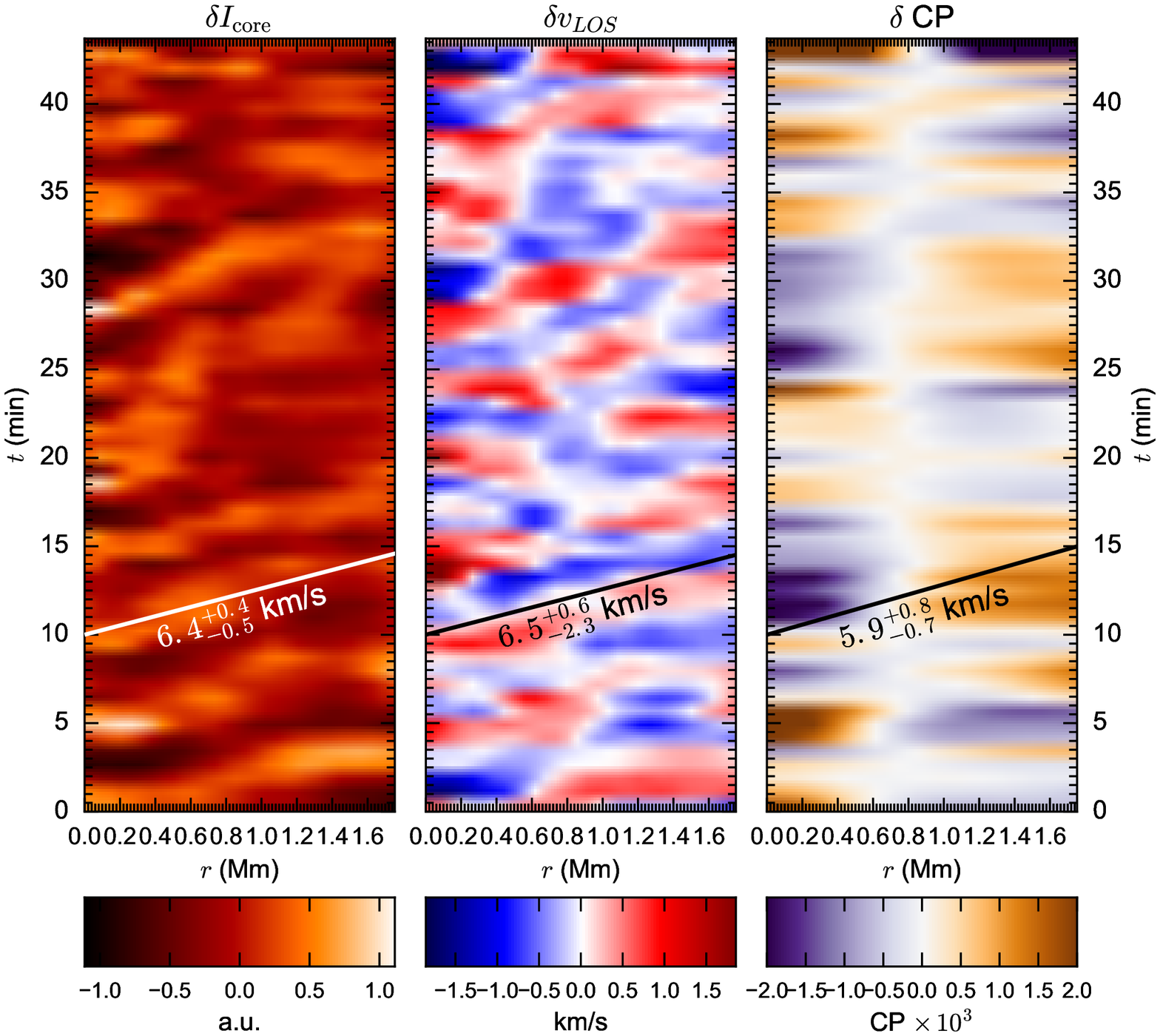}
 \includegraphics[scale=0.42,clip, trim=0 0 0 0]{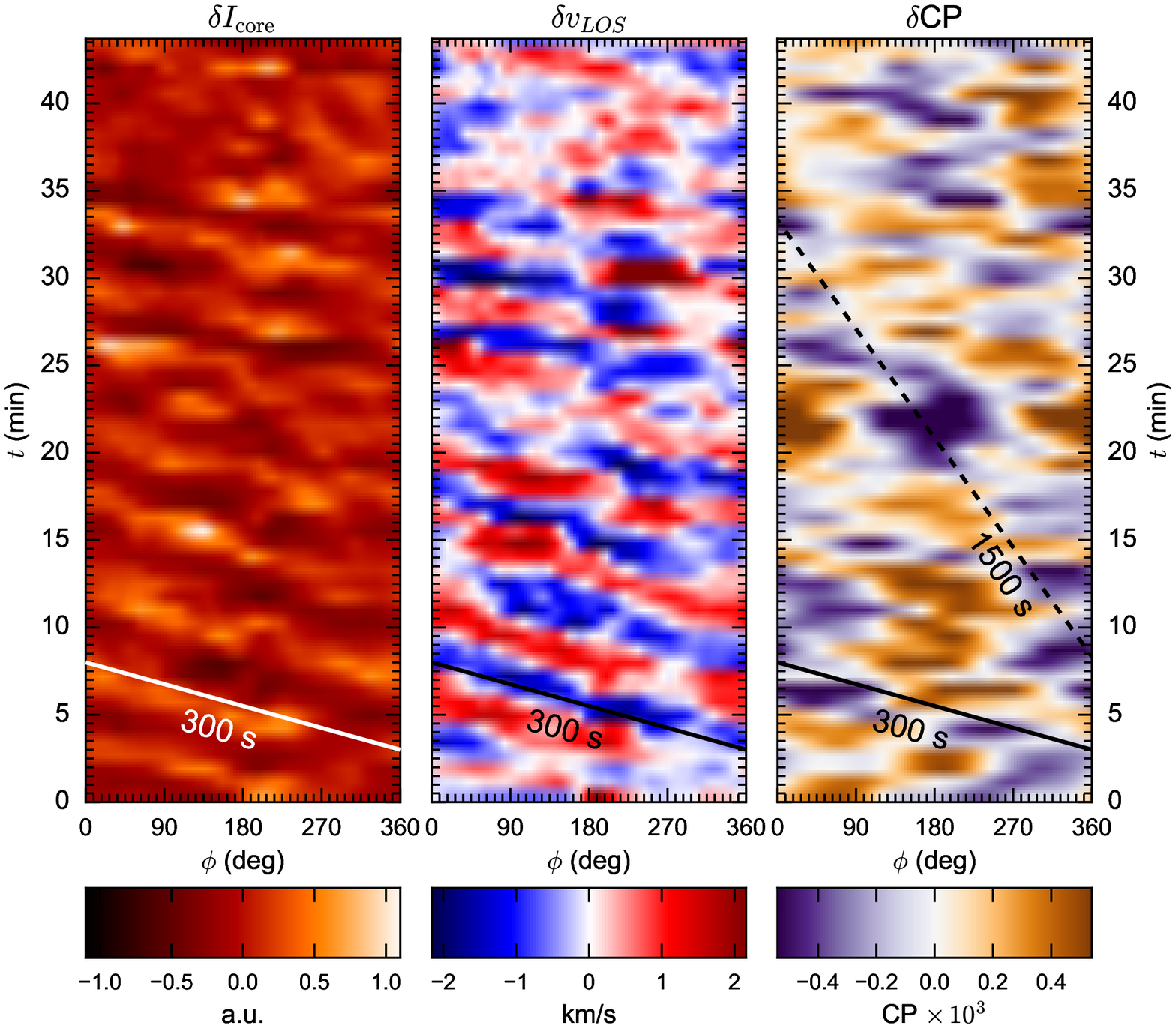}
    \caption{Top panel: Perturbations as a function of radial distance and time in intensity (left), LOS Doppler velocity (middle) and CP (right) in Ca II in a polar grid centered on N and averaged over azimuthal angle. The line highlights the phase propagation of the five-minute wave pattern with the speed from lag analysis. 
    Bottom panel: Perturbations in intensity (left), LOS Doppler velocity (middle) and CP (right) as a function of angle and time in a polar grid centered on S and averaged over radius. The line highlights the phase propagation of the wave patterns with their corresponding periodicity. (A movie of the data filtered around 5 minutes is available in the online material.)  \label{fig:erwin2} }
\end{figure}

To reveal the dynamics in each structure we construct from the polar maps time-distance plots for Ca II core intensity, LOS velocity and CP (see the online movie of the ROI that has been bandpass filtered in time around 5 minutes). We average over one spatial coordinate consistent with the mode propagation symmetry found in the spectral analysis in each structure. An overall mean is then subtracted to produce a perturbed signal. The CP perturbation is a proxy for the LOS magnetic field and magnetic pressure perturbations. The results are shown in Figure~\ref{fig:erwin2}. Following the spectral analysis, we average the signal in N across angle $\phi$ to focus on the wave propagation as a function of $r$ and time. The radially propagating waves are clear in intensity and LOS velocity, but not immediately visible in CP. We determine the speed of propagation from phase lag comparison between different radii and find it to be consistent with the spectral analysis.


For S we average the signal across radius (up to 1.7 Mm) to examine wave propagation as a function of $\phi$ and time. The clockwise propagating five-minute mode with an $|m|$=1 azimuthal structure is visible in all signatures. At a radius of 1 Mm the phase speed reaches 21 km s$^{-1}$. Furthermore, cross-wavelet analysis, as used in \citet{Shetye2019}, reveals that the intensity perturbation leads the LOS velocity perturbation by a phase shift of approximately 90 degrees, and is approximately in anti-phase with the CP perturbation. The signatures and phase relations are consistent with slow magnetoacoustic kink waves. The perturbations in CP may also be consistent with effects from opacity. Indeed, in the presence of a height dependence of the magnetic field (i.e. magnetic expansion), any fluctuation of the height of formation of the line may lead to spurious magnetic fluctuations which should not be considered as real magnetic oscillations. The CP contains variations with a $|m|$=1 structure on a longer time-scale of approximately 25 minutes. The slow rotation deduced from the optical flow analysis is not immediately visible here due to the dominance of the five-minute waves. But it is recovered if the waves are filtered out by smoothing over a longer time-scale of about 10-15 minutes.

\section{Discussion and Conclusion} \label{sec:dis}

We have reported unique spectropolarimetric measurements 
of two quiet-Sun chromospheric structures of opposite magnetic polarity, N and S, showing clockwise rotation. These represent the first polarimetric signals of vortices in the chromosphere, and confirm what was previously only inferred  indirectly. We reveal a clear pattern of circular polarization signal in and around these two structures across the whole observational period.  Although the linear polarization signal at each single time step is weak, the time-integrated signal is well above the noise. The combined measurements of circular and linear polarisations patterns provide insight into the LOS and horizontal components of the magnetic field at the chromospheric level. In summary, polarimetry reveals the presence of two regions of opposite LOS magnetic field that are separated by a reverse S-shaped polarity inversion line where the magnetic field is predominantly horizontal. The negative LOS magnetic flux region partially wraps itself around the positive flux region and S vortex. This pattern could be consistent with a non-potential magnetic dipole where both structures are magnetically connected. There is also a suggestion of a plasma flow from N towards and around S that is co-spatial with the magnetic polarity inversion line. It is worth of noting that the global picture of the N and S linked structures in  our observations resembles highly twisted and sheared sigmoid magnetic structures in the coronal active regions \citep{Canfield1999}. Such structures are usually linked to rotation in sunspots \citep{Tian2006} indicating sheared and twisted magnetic field configurations \citep[e.g.,][]{Jiang2014,Romano2014}, but also in rotating network magnetic fields \citep{Yan2015}.

\begin{figure}[ht]
    \centering
    \includegraphics[scale=0.3, clip, trim=0 0 40 0]{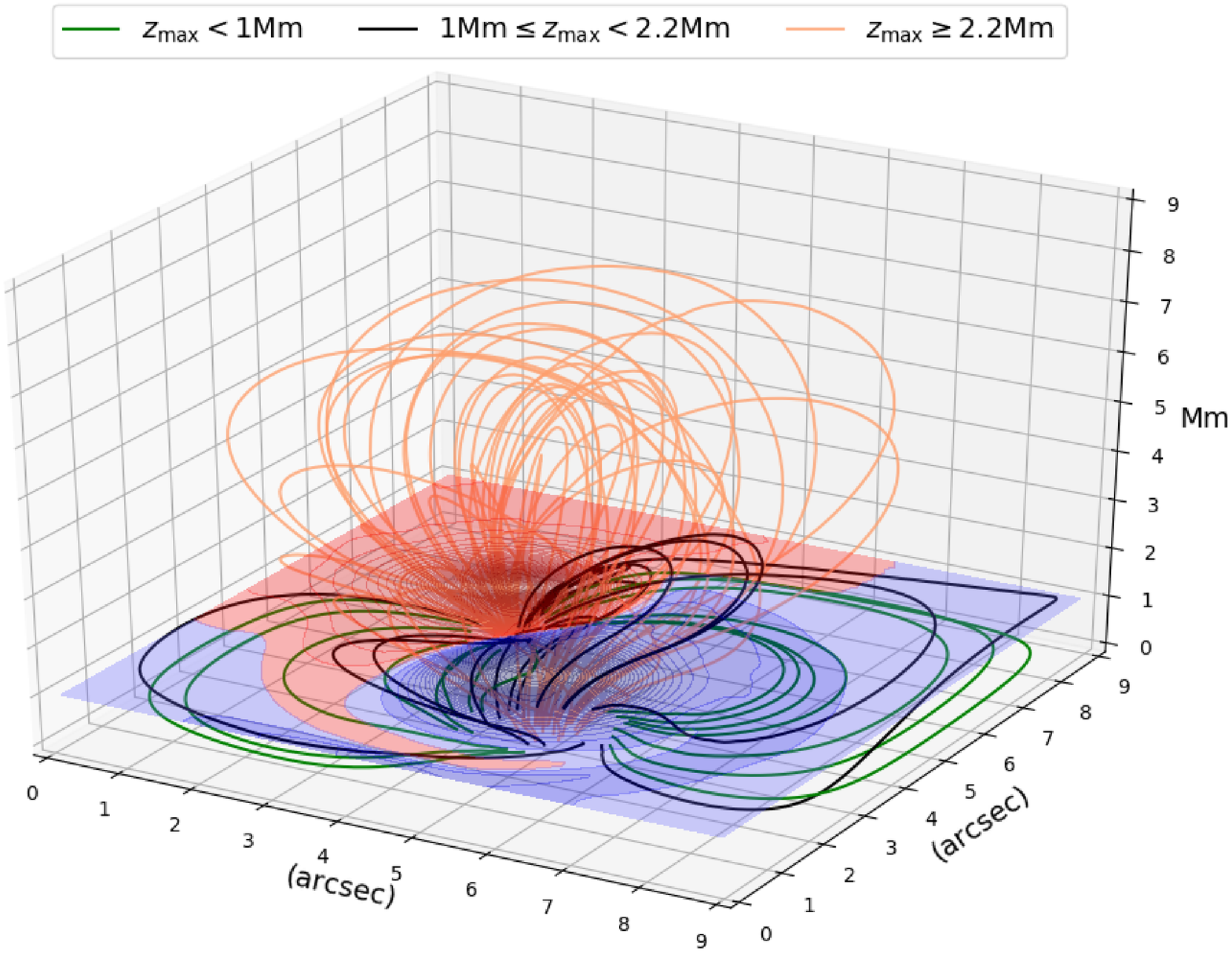}
        \includegraphics[scale=0.3, clip, trim=0 0 50 0]{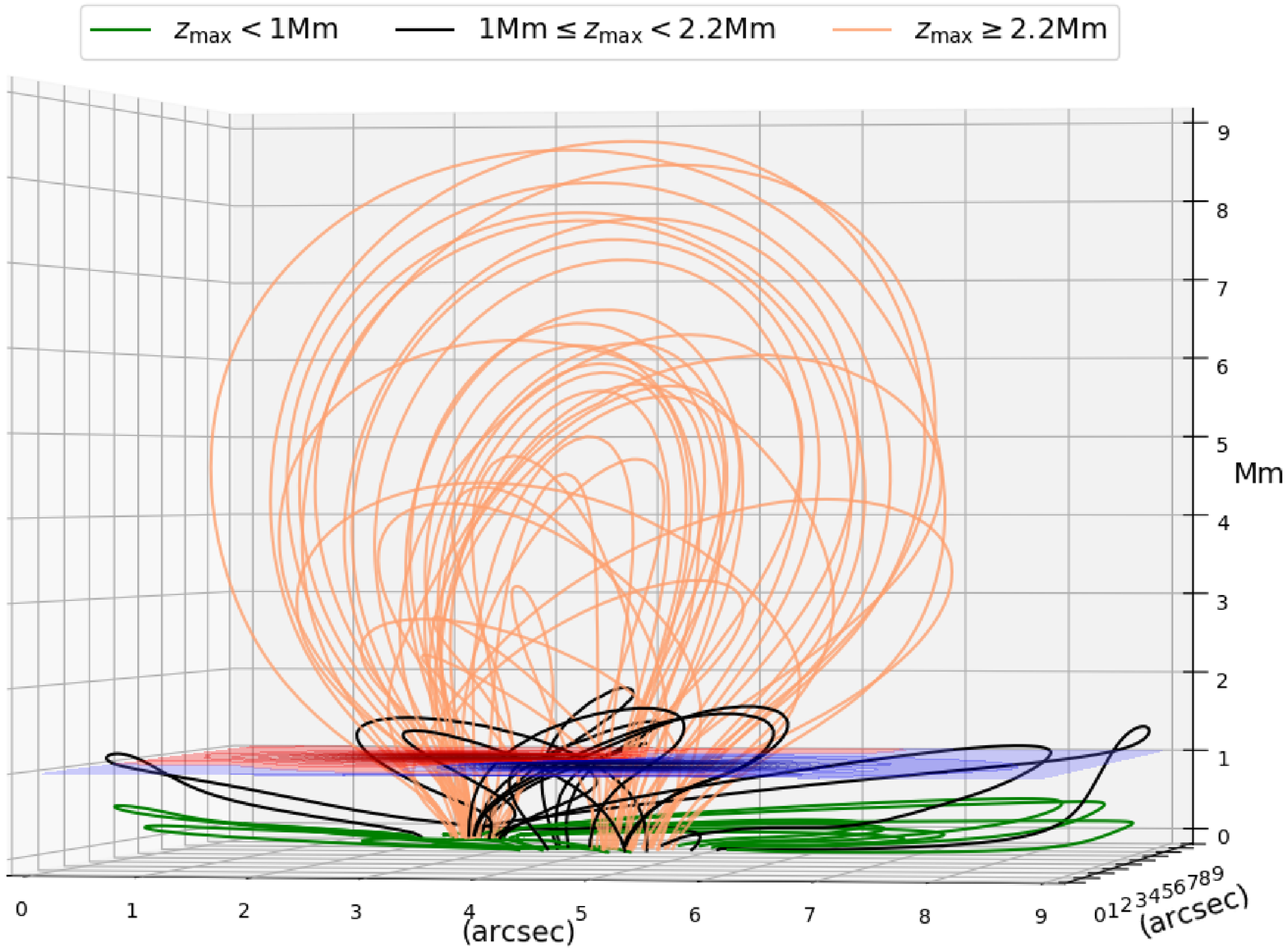}
   \caption{3D magnetic field configuration inferred from the 3D MHD simulation. The red and blue filled contours show the LOS magnetic flux at chromospheric height with red (blue) representing positive (negative) flux. A selection of magnetic field lines are traced.
   \label{fig:a2}}
\end{figure}

We reconstruct a candidate magnetic configuration of the two structures by using the Lare3d MHD code, which solves the non-linear MHD equations in three spatial dimensions \citep[][]{Lare3d}\footnote{The current version of the code is available at the following link : https://warwick.ac.uk/fac/sci/physics/research/cfsa/people/tda/larexd/}. The configuration was initialized with the VAL-C \citep{Valc} model atmosphere and a potential field. The exact location of the S vortex flux is uncertain. 

However, the observed photospheric CP shows a clear evidence of a flux concentration below the N, but no clear flux concentration below S.  The chromospheric CP shows a clear dipolar nature. We assume that the photospheric source of the S vortex is composed of diffuse weak field. Therefore, we set up the simulations with a photospheric source of 1 kG negative field with 0.5 Mm radius at N, and a larger area of weaker positive field – taken as 0.25 kG with a radius of 1 Mm at S. The S flux source is then rotated slowly by 360 degrees.

Figure \ref{fig:a2} shows the converged three-dimensional field.
It exhibits many features that are qualitatively in common with the spectropolarimetric observations presented above, i.e. distinctly shaped flux regions in the chromosphere that includes wrapping of positive flux around the S vortex and a reverse S-shaped polarity inversion line with predominantly horizontal magnetic field between N and S sources. The horizontal field lines are co-spatial with the flow channel determined from the velocity analysis. The simulation underpins our conclusion that the two structures may be magnetically connected and that through their rotation this non-potential configuration is achieved. This simulation set up will form the basis for future follow-up studies investigating rotation, flow and wave propagation.

Previous studies have treated solar vortices in isolation or at least composed of individual substructures that form and reappear in and around vortex flow \citep{Tziotziou2018}. Opposite polarity flux sources associated with a vortex are often found close to each other \citep{Wedemeyer2012,Shetye2019,Jiajia2019}. Our observations show two clockwise rotating structures of opposite magnetic polarity over a field of view of only 10 arcsec$^{2}$. Our analysis of the spectropolarimetric signatures, flow patterns and MHD simulation have shown that the two structures can be magnetically connected locally through the chromosphere. This is of consequence for quantifying the transport of mass and energy into the corona through vortices. 

The available observations do not allow us to investigate the formation of the two rotating structures as they appear to be already formed and stable at the beginning of observation time.

The two structures rotate clockwise over the whole observational period, with lower rotation rates than those previously reported for chromospheric vortices \citep{Chroswirls,Tziotziou2018}, but with a speed consistent with the rotation rates observed in photospheric vortices \citep{Bonet2010,Vargas2011}. Near their centres the two structures rotate as a rigid body.
 
Intensity, Doppler velocity and CP in the chromosphere reveal five-minute waves, similar to those reported from analysis of vortex motion by \citet{Shetye2019} and \citet{Tziotziou2019}. At each structure these waves exhibit different local propagation characteristics. We interpret these local waves as slow magnetoacoustic in nature because their periodicity matches that found dominating throughout the FOV, have significant LOS velocity and intensity signatures, consistent phase relations and lack oscillatory behaviour in the azimuthal flow. The N structure carries (in projection) radially outward propagating waves. The waves in the S vortex propagate azimuthally with an $m$=1 phase speed pattern. Resonant absorption of sound waves by a magnetic flux tube favours the $|m|$=1 mode if the tube is thinner than the wavelength  \citep{Sakurai1991}. The preference for clockwise propagation is indicative of an azimuthal asymmetry such as magnetic twist or rotation.  

We infer that the clock-wise rotational S vortex pattern is due to a combination of rotational mass flow up to 1.5 km s$^{-1}$ and an |m|=1 slow magnetoacoustic mode with a period of approximately 300 s and a phase speed of 20 km s$^{-1}$ at the vortex edge. The flow would cause a Doppler shift in the mode of about 25s but this cannot be discerned with the time cadence of 46s.

In general, chromospheric vortices may exhibit a superposition of fast rotational phase patterns due to MHD waves on top of a slower motion due to actual rotation \citep{Tziotziou2019, Shetye2019}.
Careful wave-analysis will be essential to distinguish steady rotation, wave phase speed and wave amplitude to obtain accurate measurements of the Poynting flux into the solar corona from MHD waves associated with vortices.

A comprehensive survey of the overall magnetic field topology of solar vortices is needed to assess the channelling of energy to higher atmospheric layers. 
Chromospheric spectropolarimetric measurements will be essential. In this regard,  
4-meter class solar telescopes will enable new higher sensitivity and accuracy spectropolarimetric measurements than those analysed in the present study to possibly provide further observational evidence of these processes down to unprecedented spatial resolutions.

\begin{acknowledgements}
We thank the referee for the useful suggestions. The authors thank Doug Gilliam (NSO) and Michiel Van  Noort  (MPS)  for  their  valuable  support  during the  acquisition  and  MOMFBD  processing  of  the  data. The  research  leading  to  these  results  has received  funding from the European Union's Horizon 2020 research and innovation programme under Grant Agreement No 824135 (SOLARNET), the Istituto Nazionale di Astrofisica (INAF) and the Italian MIUR-PRIN grant 2017 ''Circumterrestrial Environment: Impact of Sun$–$Earth Interaction''.  JS is funded by STFC Grant ST/P000320/1. JS would like to thank INAF for supported visits. This research has made use of the IBIS-A archive.
MS would like to thank University of Warwick for supported visit. The authors wish to acknowledge scientific discussions with the Waves in the Lower Solar Atmosphere (WaLSA; www.WaLSA.team) team, which is supported by the Research Council of Norway (project number 262622), and The Royal Society through the award of funding to host the Theo Murphy Discussion Meeting “High resolution wave dynamics in the lower solar atmosphere” (grant Hooke18b/SCTM).

\end{acknowledgements}

%

\begin{thebibliography}{}


      
      
\bibitem[Arber et al. (2012)]{Lare3d} Arber, T. D., Longbottom, A. W., Gerrard, et al.\ 2012, Astrophysics Source Code Library

\bibitem[Attie et al.(2009)]{Attie2009} Attie, R., Innes, D.~E., \& Potts, H.~E.\ 2009, \aap, 493, L13

\bibitem[Bogdan(1989)]{Bogdan1989} Bogdan, T.~J.\ 1989, \apj, 339, 1132

\bibitem[Bonet et al.(2010)]{Bonet2010} Bonet, J.~A., M{\'a}rquez, I., S{\'a}nchez Almeida, J., et al.\ 2010, \apj, 723, L139

\bibitem[Canfield et al.(1999)]{Canfield1999} Canfield, R.~C., Hudson, H.~S., \& McKenzie, D.~E.\ 1999, \grl, 26, 627

\bibitem[Cavallini(2006)]{Cav06} Cavallini, F. 2006, Sol. Phys., 236, 415

\bibitem[Farneback (2003)]{Farneback} Farneback, G\ 2003, Scandinavian Conference on Image Analysis, 2749, 363

\bibitem[Fedun et al.(2011)]{Fedun2011} Fedun, V., Shelyag, S., Verth, G., et al.\ 2011, Annales Geophysicae, 29, 1029

\bibitem[Hale(1908)]{Haleb} Hale, G.~E.\ 1908, \apj, 28, 100

\bibitem[Kitiashvili et al.(2013)]{Kitiashvili2013} Kitiashvili, I.~N., Kosovichev, A.~G., Lele, S.~K., et al.\ 2013, \apj, 770, 37

\bibitem[Jiang et al.(2014)]{Jiang2014} Jiang, C., Wu, S.~T., Feng, X., et al.\ 2014, \apj, 780, 55

\bibitem[L\"ofdahl(2002)]{Lof02} L\"ofdahl, M.G. 2002, SPIE, 4792, 146LS

\bibitem[Liu et al.(2019a)]{Jiajia2019} Liu, J., Carlsson, M., Nelson, C.~J., et al.\ 2019, \aap, 632, A97

\bibitem[Liu et al.(2019b)]{Liu2019b} Liu, J., Nelson, C.~J., \& Erd{\'e}lyi, R.\ 2019, \apj, 872, 22

\bibitem[Liu et al.(2019c)]{Liu2019c} Liu, J., Nelson, C.~J., Snow, B., et al.\ 2019, Nature Communications, 10, 3504

\bibitem[Mart{\'\i}nez Pillet et al.(2011)]{Pillet2011} Mart{\'\i}nez Pillet, V., Del Toro Iniesta, J.~C., {\'A}lvarez-Herrero, A., et al.\ 2011, \solphys, 268, 57

\bibitem[Moll et al.(2011)]{Moll2011} Moll, R., Cameron, R.~H., \& Sch{\"u}ssler, M.\ 2011, \aap, 533, A126

\bibitem[Quintero Noda et al.(2016)]{QuinteroNoda16} Quintero Noda, C., Shimizu, T., de la Cruz Rodr{\'\i}guez, J., et al.\ 2016, \mnras, 459, 3363

\bibitem[Requerey et al.(2018)]{Requerey2018} Requerey, I.~S., Cobo, B.~R., Go{\v s}i{\'c}, M., \& Bellot Rubio, L.~R.\ 2018, \aap, 610, A84 

\bibitem[Romano et al.(2014)]{Romano2014} Romano, P., Zuccarello, F.~P., Guglielmino, S.~L., et al.\ 2014, \apj, 794, 118

\bibitem[Sakurai et al.(1991)]{Sakurai1991} Sakurai, T., Goossens, M., \& Hollweg, J.~V.\ 1991, \solphys, 133, 247 

\bibitem[Schmieder et al.(2017)]{Schmieder2017} Schmieder, B., Mein, P., Mein, N., et al.\ 2017, \aap, 597, A109 

\bibitem[Shetye et al. (2019)]{Shetye2019} Shetye, J., Verwichte, E., Stangalini, M., et al.\ 2019, \apj, 881, 83

\bibitem[Tian \& Alexander(2006)]{Tian2006} Tian, L., \& Alexander, D.\ 2006, \solphys, 233, 29

\bibitem[Tziotziou et al.(2018)]{Tziotziou2018} Tziotziou, K., Tsiropoula, G., Kontogiannis, I., Scullion, E., \& Doyle, J.~G.\ 2018, \aap, 618, A51

\bibitem[Tziotziou et al.(2019)]{Tziotziou2019} Tziotziou, K., Tsiropoula, G., \& Kontogiannis, I.\ 2019, \aap, 623, A160

\bibitem[Vargas Dom{\'\i}nguez et al.(2011)]{Vargas2011} Vargas Dom{\'\i}nguez, S., Palacios, J., Balmaceda, L., et al.\ 2011, \mnras, 416, 148

\bibitem[Vernazza et al.(1981)]{Valc} Vernazza, J. E., Avrett, E. H., Loeser, R.\ 1981, \apj Suppl. Ser., Vol. 45, p. 635-725

\bibitem[Wedemeyer-B{\"o}hm \&  Rouppe van der Voort(2009)]{Chroswirls} Wedemeyer-B{\"o}hm, S., \& Rouppe van der Voort, L.\ 2009, \aap, 507, L9 

\bibitem[Wedemeyer-B{\"o}hm et al.(2012)]{Wedemeyer2012} Wedemeyer-B{\"o}hm, S., Scullion, E., Steiner, O., et al.\ 2012, \nat, 486, 505 

\bibitem[Wedemeyer et al.(2013)]{Wedemeyer2013} Wedemeyer, S., Scullion, E., Rouppe van der Voort, L., Bosnjak, A., \& Antolin, P.\ 2013, \apj, 774, 123

\bibitem[Wedemeyer \& Steiner(2014)]{Wedemeyer2014} Wedemeyer, S., \& Steiner, O.\ 2014, \pasj, 66, S10

\bibitem[Shelyag et al.(2013)]{Shelyag2013} Shelyag, S., Cally, P.~S., Reid, A., et al.\ 2013, \apjl, 776, L4

\bibitem[Yadav et al.(2020)]{Yadav2020} Yadav, N., Cameron, R.~H., \& Solanki, S.~K.\ 2020, arXiv e-prints, arXiv:2004.13996

\bibitem[Yan et al.(2015)]{Yan2015} Yan, X.~L., Xue, Z.~K., Pan, G.~M., et al.\ 2015, \apjs, 219, 17
 



\end{thebibliography}
%

\end{document}